\documentclass[epjCONF]{svjour}
\usepackage{graphicx}
\usepackage[varg]{txfonts} 
\usepackage[latin1]{inputenc}
\usepackage{amstext}
\usepackage{array}
\session-title{MESON2012 - 12th International Workshop on Meson Production, Properties and Interaction}
\begin{document}
\newcolumntype{C}{>{$}c<{$}}
\newcolumntype{R}{>{$}r<{$}}
\newcolumntype{L}{>{$}l<{$}}
\title{Bayesian inference of the resonance content of $p(\gamma,K^+)\Lambda$}
\author{P.~Vancraeyveld\thanks{\email{pieter.vancraeyveld@ugent.be}} \and 
  L.~De~Cruz \and
  J.~Ryckebusch\thanks{\email{jan.ryckebusch@ugent.be}} \and
  T.~Vrancx
}
\institute{Department of Physics and Astronomy, Ghent University, Proeftuinstraat 86, B-9000, Gent, Belgium, {http://inwpent5.ugent.be}}
\abstract{
A Bayesian analysis of the world's $p(\gamma,K^+)\Lambda$ data is presented.
We adopt a Regge-plus-resonance framework featuring consistent couplings for nucleon resonances up to spin $J=5/2$,
and evaluate 2048 model variants considering all possible combinations of 11 candidate resonances.
The best model, labeled RPR-2011, is discussed with special emphasis on nucleon resonances in the 1900-MeV mass region. 
} 
\maketitle
\section{Introduction}
\label{sec:intro}

Dedicated experiments have lead to an impressive increase in the quantity and quality
of e\-lec\-tro\-mag\-net\-ic-meson-production data.
This situation is likely to improve even further in years to come.
With such an abundance of experimental information,
one anticipates great progress in our understanding of the nucleon and its excited states.
In order to establish the resonance content of meson-production reactions,
a number of challenges need to be dealt with:
couplings between different reaction channels,
the occurrence of broad and overlapping resonances,
and under-determined data sets that do not allow to establish a unique best-fit model.
This state of affairs prompts for a further refinement of model ingredients,
but also calls for a re-evaluation of the machinery used for model selection.

Electromagnetic production of hyperons~($Y$) and kaons~($K$) plays a central role
in the search for hitherto unidentified resonances that have been predicted by quark models.
From an experimentalist's perspective,
kaon production is appealing,
despite its small cross section,
because the self-analyzing power of hyperons facilitates the measurement of polarization observables.
The CLAS collaboration, for instance,
have recently collected data that constitutes an over-determined experiment~\cite{CLASproceedings}.

Because of the relatively high threshold for $KY$ production,
the reaction mechanism has contributions from a large number of broad and overlapping nucleon resonances~($N^\ast$).
In addition, a crucial role is reserved for non-resonant diagrams.
This observation constitutes a challenge for single- and coupled-channel partial-wave analyses,
as well as traditional isobar models;
the resonance content of $N(\gamma,K)Y$ can only be determined
after resonant and non-resonant contributions are cleanly separated.
The Regge-plus-resonance~(RPR) formalism elegantly addresses this issue
via a hybrid approach combining a hadrodynamic description in the resonance region
with Regge phenomenology constrained at higher energies.

In this contribution,
we discuss a Bayesian analysis of the world's $p(\gamma,K^+)\Lambda$ data adopting the RPR framework.
Section~\ref{sec:formalism} summarizes the RPR formalism,
and introduces Bayesian-evidence computation as a powerful tool for model selection.
In Section~\ref{sec:results},
the optimal resonance content for $p(\gamma,K^+)\Lambda$ is discussed
with special emphasis on the 1900-MeV mass region.
Finally, we present our conclusions.

\section{The Regge-plus-resonance framework and Bayesian model selection}
\label{sec:formalism}

In the RPR approach~\cite{RPRlambda},
the non-resonant contributions to $p(\gamma,K^+)\Lambda$ are parametrized in terms of
$K^+(494)$ and $K^{\ast+}(892)$ Regge-trajectory $t$-channel exchange.
To ensure gauge invariance,
the electric part of a Reggeized $s$-channel Born diagram is added.
This results in a three-parameter model that successfully reproduces data at high energies,
and describes the general features of $p(\gamma,K^+)\Lambda$ in the resonance region.

The model is refined
by considering contributions from $s$-channel $N^\ast$ exchanges using standard Feynman diagrams.
A key aspect of the current analysis is the use of consistent couplings~\cite{RPRconsistent}
for high-spin ($J\geq3/2$) resonance exchange.
This formalism dictates that each spin-1/2 resonance-exchange diagram introduces one parameter;
high-spin resonances require two parameters.
Moreover,
on account of the adopted multidipole-Gauss form factor~\cite{RPRconsistent},
a single cut-off parameter suffices in the hadronic form factors of the $N^\ast$-exchange diagrams.
As a result,
the RPR framework allows for an unambiguous separation of resonant and non-resonant diagrams
while the number of free parameters is kept to a bare minimum.
The latter is important in the light of Bayesian model selection.

As mentioned in the previous section,
the tools used for model selection are crucial to establish the resonance content of meson-production reactions.
For a model with a set of parameters $\alpha_M$ one conventionally defines
\begin{equation}\label{eq:chi2}
\chi^2(\alpha_M) = \sum_{i=1}^N \frac{\left[d_i-f_i(\alpha_M)\right]^2}{\sigma_i^2}\,,
\end{equation}
where $N$ is the total number of data points,
$\sigma_i$ is the error on data point $d_i$,
and $f_i(\alpha_M)$ is the corresponding model prediction.
The quantity in Eq.~\ref{eq:chi2} obeys a chi-square distribution,
and parametrizes the likelihood function $\mathcal{L}(\alpha_M)$.
A common approach to model selection involves $\chi^2$ minimization (equivalent to likelihood maximization)
for different model variants,
and subsequent comparison of the minimal $\chi^2$ values.
A major drawback of this method is the disregard of Occam's razor.
Many recipes exist to quantitatively penalize the introduction of superfluous model parameters.
They are, however, not statistically rigorous.

An alternative procedure, which we have adopted, is offered by the Bayesian evidence
\begin{equation}
\mathcal{Z} = \int \mathcal{L}(\alpha_M)\pi(\alpha_M)d\alpha_M\,,
\end{equation}
where the prior distribution $\pi(\alpha_M)$ expresses any prior knowledge of the parameters' probability distribution.
The numerical computation of $\mathcal{Z}$ is highly nontrivial,
and requires a dedicated set of algorithms~\cite{RPR2011}.
Evaluating the Bayesian evidence for different models $M_A$ and $M_B$
allows to quantitatively express the relative probability of both models
given the available experimental data $\left\lbrace d_k \right\rbrace$~\cite{RPRprl,RPR2011}
\begin{equation}
\frac{\mathcal{Z}_A}{\mathcal{Z}_B} = 
 \frac{P(M_A|\left\lbrace d_k \right\rbrace)}{P(M_B|\left\lbrace d_k \right\rbrace)} \,.
\end{equation}
The natural logarithm of the evidence ratio can be interpreted qualitatively with the aid of Jeffreys' scale,
given in Table I of Ref.~\cite{RPR2011}.
To appropriately handle the statistical and systematic errors from various experiments,
a rescaled Bayesian evidence $\mathcal{Z}^\prime$ is introduced~\cite{RPRprl,RPR2011}.

\section{Resonance content of $p(\gamma,K^+)\Lambda$ in the RPR-2011 model}
\label{sec:results}

\begin{figure}
\centering
\includegraphics[width=.7\textwidth]{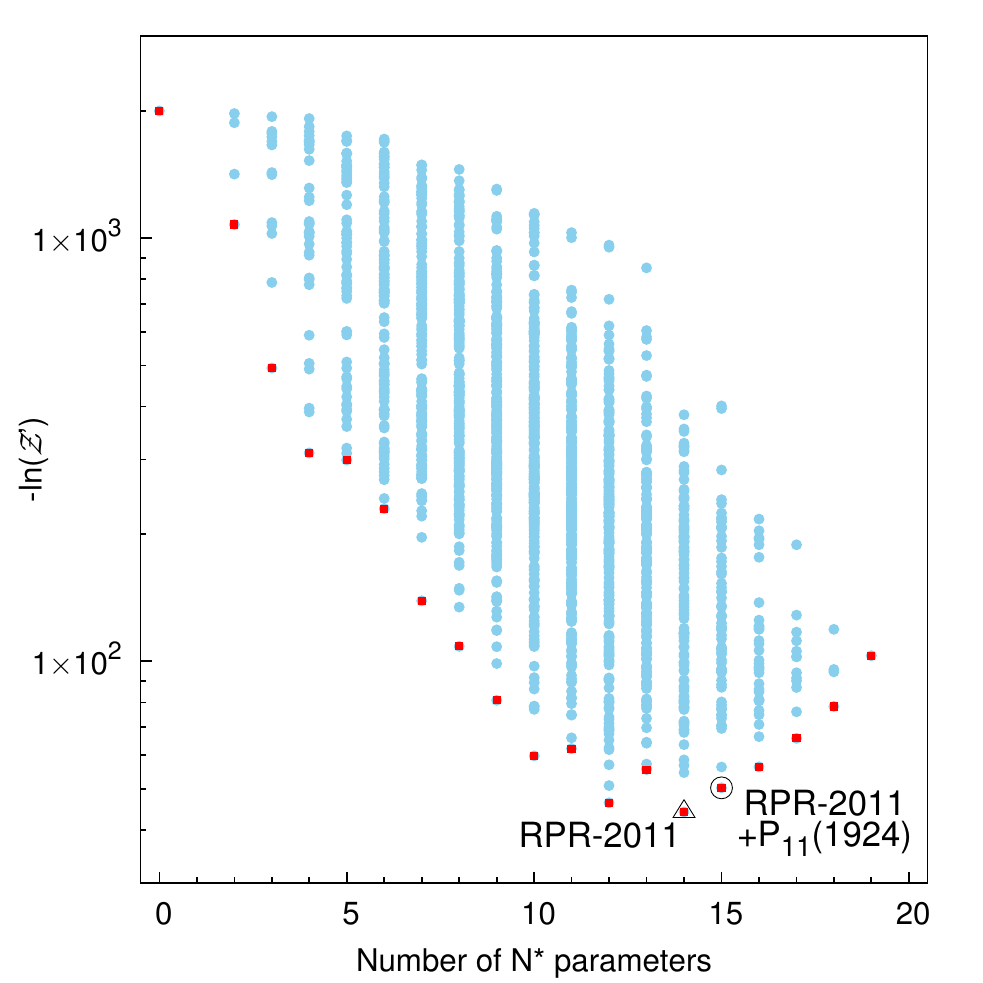}\caption{%
The evidences $(-\ln\mathcal{Z}^\prime)$ of the 2048+1 model variants in the RPR model space (blue circles, see text),
as a function of the number of $N^\ast$ parameters.
The best model for a fixed number of parameters is indicated with a red square.
The RPR-2011 model, which has the highest evidence, is marked with an open triangle.
The open circle denotes the RPR-2011 model supplemented with the $P_{11}(1924)$ resonance.}
\label{fig:evidence}       
\end{figure}

In Refs.~\cite{RPRprl,RPR2011},
we presented the details of a Bayesian analysis of the world's $p(\gamma,K^+)\Lambda$ data in the RPR approach
as outlined in the previous section.
Fig.~\ref{fig:evidence} plots
the computed evidence values for 2048 model variants
constructed by combining the 11 candidate nucleon resonances with spin $J\leq5/2$ listed in Table V of Ref.~\cite{RPR2011}.
The model with the highest evidence, referred to as RPR-2011,
has 14 $N^\ast$ parameters and features the 
$S_{11}(1535)$, $S_{11}(1650)$, $F_{15}(1680)$, $P_{13}(1720)$,
$P_{13}(1900)$, $F_{15}(2000)$, and the missing $D_{13}(1900)$ and
$P_{11}(1900)$ resonances~\cite{RPRprl}.

While the existence of some resonances, e.g.\ $S_{11}(1650)$ and $P_{13}(1720)$,
is confirmed in various reaction channels by many models,
the status of resonances in the 1900-MeV mass range is unsettled.
An overview of considered $N^\ast$s in this mass range is given in Table~\ref{tab:resonances}.
Using Bayesian inference, we find decisive evidence for the $P_{13}(1900)$ state 
that was promoted from two- to three-star status in the latest PDG update~\cite{pdg}.
In addition,
RPR-2011 boasts two states labeled as ``missing''.
Indeed, we find for the $P_{11}(1900)$ and $D_{13}(1900)$
decisive and significant to strong evidence respectively~\cite{RPR2011}.

The presence of a resonance with $J^\pi=1/2^+$ and a mass close to 1900 MeV,
might point towards the contribution of a
bound state of the $K\bar{K}N$ system first found in Ref.~\cite{JidoOsetResonancePrediction}.
This $P_{11}(1924)$ resonance was confirmed in other model calculations~\cite{OsetResonancePrediction},
and predicted to have a significant decay in the $K\Lambda$-production channel~\cite{OsetResonanceKL}.
The Bayesian methodology that we adopt allows to quantitatively express 
the conditional probability $P(R|D)$ ($P(\neg R|D)$) that a resonance $R$ is (not) needed, given experimental data $D$.
Evaluating the probability ratio $P(R|D)/P(\neg R|D)$ for the $P_{11}(1924)$ resonance given the present data
would require us to calculate the Bayesian evidence $\mathcal{Z}^\prime$ of an additional 2048 model variants
-- a CPU-time consuming endeavor.
In order to get an impression of a possible role of this resonance,
we have limited ourselves to the evaluation of $\mathcal{Z}^\prime$ for the RPR-2011 set of resonances
supplemented with the $P_{11}(1924)$.
This model variant encompasses 15 $N^\ast$ parameters,
and is marked in Fig.~\ref{fig:evidence} with an open circle.
Adding the $P_{11}(1924)$ state to the RPR-2011 set results in the best 15-parameter model,
but does not surpass the evidence of RPR-2011.
We find $\Delta\ln\mathcal{Z}^\prime = 6.17$,
and thus decisive evidence in favor of the RPR-2011 model according to Jeffreys' scale.
This finding does not exclude a role for the $P_{11}(1924)$ resonance in $p(\gamma,K^+)\Lambda$,
since an alternative set of resonances including the $P_{11}(1924)$ might outperform RPR-2011.

\begin{table}
\caption{%
The considered nucleon resonances in the $1900\,\text{MeV}$ mass range
given in the notation $L_{2I,2J}(M)$,
along with the PDG status, spin-parity ($J^\pi$),
Breit-Wigner mass ($M$) and width ($\Gamma$).
}
\label{tab:resonances}       
\centering
\begin{tabular}{LCCCC}
\hline\noalign{\smallskip}
\text{Resonance} & \text{PDG status} &J^\pi & M\,(\text{MeV}) &\Gamma\,(\text{MeV})  \\
\noalign{\smallskip}\hline\noalign{\smallskip}
D_{13}(1900) &\text{missing} &3/2^- &1895 &200 \\
P_{11}(1900) &\text{missing} &1/2^+ &1895 &200 \\
P_{13}(1900) &{\ast}{\ast}{\ast} &3/2^+ &1900 &500^{+80}_{-360} \\
P_{11}(1924) &\text{missing} &1/2^+ &1924 &60 \\
\noalign{\smallskip}\hline
\end{tabular}
\end{table}

\section{Conclusions}

The RPR framework merges Reggeized $t$-channel exchange with
tree-level $s$-channel $N^{\ast}$-exchange contributions from an isobar approach
into an economical model for kaon photoproduction in and above the resonance region. 
The RPR formalism clearly separates non-resonant from resonant amplitudes
which is an asset when searching for the properties of resonances which contribute to $p(\gamma,K^+)\Lambda$. 
We have used Bayesian inference to perform model selection.  

Starting from a set of eleven candidate resonances,
the optimum model for $p(\gamma,K^+)\Lambda$, dubbed \mbox{RPR-2011}, is determined.
This model encompasses the resonances $S_{11}(1535)$,
$S_{11}(1650)$, $F_{15}(1680)$, $P_{13}(1720)$, $P_{11}(1900)$,
$F_{15}(2000)$, $D_{13}(1900)$, and $P_{13}(1900)$.
Decisive and significant to strong evidence is found for two states in the 1900-MeV mass range
that are not listed in the PDG.
Finally,
we have explored the possible role for a $K\bar{K}N$ bound state with $J^\pi=1/2^+$.
Our initial investigation,
whereby the $P_{11}(1924)$ resonance is added to the RPR-2011 set,
has resulted in a 15-parameter model with the third highest evidence of all 2048+1 considered models.
Nonetheless, given the current data
our analysis indicates decisive evidence in favor of the $N^\ast$ combination of the RPR-2011 model.

\begin{acknowledgement}
This research was financed by the Flemish Research Foundation (FWO Vlaanderen),
and the research council of Ghent University.
The computational resources (Stevin Supercomputer Infrastructure) and services used in this work were provided 
by Ghent University, the Hercules Foundation and the Flemish Government -- department EWI.
\end{acknowledgement}

\bibliographystyle{epj}
\bibliography{bibliography}

\end{document}